\documentstyle[aps,prd,twocolumn]{revtex}

\newcommand{\rf}[1]{(\ref{eq:#1})}

\begin{document}

\title{ Geometric Stability of Brane-worlds} 
\author{ M. D. Maia\\
Universidade de Bras\'{\i}lia, Instituto  de F\'{\i}sica \\ 
 Bras\'{\i}lia. D.F. 7919-970\\{maia@fis.unb.br} \\ and\\
Edmundo M. Monte\\ 
Universidade Federal da Para\'{\i}ba\\
Departamento de F\'{\i}sica\\
Jo\~ao Pessoa, Pb. 58100-000\\{edmundo@fis.ufpb.br} }

\maketitle

\begin{abstract}
\begin{center}
 {\bf Abstract}
\end{center}
The stability  conditions for  coordinate gauge independent perturbations of  brane-worlds   are analyzed. It is shown that, these conditions lead to the  Einstein-Hilbert dynamics and to a confined  gauge potential, independently  of  models and   metric ansatzes.  The   size
of the extra dimensions  are estimated  without assuming a  fixed topology. 
The quantum modes  corresponding to  high frequency gravitational waves  are defined  through a canonical structure. 
\end{abstract}
\vspace{3mm}
pacs{ 11.10.Kk, 04.50.+h, 04.60.+n}

\section{ Introduction}
 The "brane-worlds" program  proposes a  unification of all interactions at  the TeV scale, with the supposition that  there are  large (as  compared to Planck's length)  extra dimensions. The hierarchy problem is  solved by the  assumption  that the four-dimensional space-time   geometry   undergoes quantum fluctuations  along the extra dimensions,  but the gauge interactions of the standard model remain  confined  within the  space-time.  This   is implemented  by the 
introduction of a   fundamental  Planck  scale in the higher dimensional bulk, while the  effective four-dimensional Planck mass is   adjusted in accordance with the geometry of the  higher dimensional space.    The geometrical scenario  compatible with these assumptions is that of   four-dimensional space-times, the  brane-worlds, are dynamically embedded in a   higher dimensional  solution Einstein's equations \cite{Arkani:1,Randall}.

In contrast  with Kaluza-Klein and Superstring theories  the brane-world unification in principle may be   experimentally  verified by high-energy collisions at   hundreds  of  TeV.  Another  implication that  may soon be checked is  the  modification of   Newton's law of   gravitation  at     sub millimeter  scale \cite{Newman}. 

There is not  yet a  theory of brane-worlds in the sense that the fundamental  principles  are  well established.  Some  earlier  publications,  not always acknowledged,  have  proposed  some  basic  concepts and  formal tools.  For example, the  confinement of 
gauge interactions to space-time  behaving as a  potential well in a higher dimensional space has been suggested  in  the early eighties   \cite{Akama:1,RS,Visser,Antoniadis}. 
 Applications of  space-time embeddings  to  the quantization of the geometry, to generate internal symmetries and  as an alternative to compactification,  have been  around for  a considerable time \cite{Friedman,RT,Deser,Maia:1,Maia:2,Davidson,Pavsic:1,Gibbons,Tapia:1,Pavsic:2,Tapia:2}. 
Recent reviews and   additional references can be found in \cite{Zurab:1,Lorenzana,Carter}.

Presently there are two  conceptually distinct  approaches to brane-worlds.  One   of them    is based on a higher dimensional space with product  topology $M_{4}\times B_{N}$,   $N\ge 2$,  where $B_{N}$ is the a finite volume internal space,  similar to Kaluza-Klein theory  \cite{Arkani:1}.
The other,  is  defined  on  a  5-dimensional  anti-deSitter geometry,  with brane-like  boundary conditions \cite{Randall}. Current problems  include the stability of the  brane-world   structure; the lack of a  proper explanation of the confinement of the gauge fields and  a  consistent definition of  the  quantum states \cite{Giddings,Akama:2}. 

The purpose of this paper is to  show that the stability of the embedding  is  a central property of brane-worlds  compatible with the aforementioned  assumptions, independently of   models and metric ansatzes. More specifically,  we   show in sections  II and III that the   stability  conditions for the  classical perturbations  of  a  brane-world, regardless of  the size,  number and signature of the extra  dimensions, provide  the   dynamics of the  brane-world evolution.  The cases  with just one extra dimensions,  studied in section IV, are found to have limited  properties, including  the need  for   confinement mechanisms which are  independent  and may interfere with the  stability conditions.
The   quantum fluctuations of the brane-worlds  described in section V,  induce   topological changes   which are  incompatible  with  a fixed  topology.  We  show that  such  topological changes  result from  the  multiparameter quantum  fluctuations. In section  VI  it is shown that the  differentiable structure of the brane-world  also implies in the existence a confined  gauge field structure.  Finally, the  concluding section   also  discusses on  the  number   and size of the  extra coordinates.

\section{ Brane-worlds perturbations}

In very general terms, a brane-world  may be  described  as  a  locally embedded  space-time, capable of  quantum fluctuations  along the extra dimensions,  but retaining the gauge interactions confined within.   These conditions  impose  dynamical  conditions on the embedding,  producing   what could be called  a {\em dynamical embedding} of space-times  as opposed to the  more common analytic embedding. 

It is well known, that any space-time   $\bar{V}_{n}$  may  be locally embedded
into a manifold  $V_{D}$, with sufficiently large dimension $D$.   
The number of extra dimensions  $N=D-n$  
depends on the  geometries of  $\bar{V}_{n}$ and of  $V_{D}$, and  of course
of the type of  embedding map, which may be  local, non-local,  isometric,  and conformal among other possibilities. 

The simplest  examples  are  those of space-times isometrically embedded in a flat space  $M_{D}$, where the  embedding is given by  analytic functions  \cite{Cartan,Janet}. The analytic assumption greatly simplifies the embedding   and it implies that   $D $ is at most $10$. 
In brane-worlds it is  not obvious  that the analyticity  condition holds under the  assumed conditions and  in special under the quantum fluctuations. However,  it seems   reasonable to assume  that the embedding  remains  differentiable, determined by the classic Gauss-Codazzi-Ricci equations. In this case,  the   limiting dimension for flat embeddings rises  to $D=14$,  with a   wide range of compatible signatures \cite{Greene}.  
We will see that the differentiable  hypothesis  is consistent with  the  Einstein-Hilbert dynamical principle,  which  lies at the basis of the mentioned  dynamical embedding.

For generality, we   consider  an n-dimensional submanifold  with  arbitrary metric signature embedded in a D-dimensional space, which is  a solution of higher dimensional Einstein's equations\footnote{The  use of  generic signature  may  cause the emergence of  ghosts. However, in view of  prospective topological  changes,  it would be unwise to any specify the signature at this stage. The expressions can be adjusted  to specific  signatures by simple arithmetics. }.

  The embedding of the background   $\bar{V}_{n}$,  with metric  $\bar{g}_{ij}$, is given by  local  the map $\bar{\cal X}: \bar{V}_{n}\rightarrow V_{D}$ such that\footnote{  
All Greek indices  run from 1 to
$D$. Small case Latin indices $i,j,k...$ run from  1  to  $n$.  An overbar
denotes an object of the background  space-time.   
The covariant derivative with respect to
the metric of  the higher dimensional manifold is denoted by  a semicolon and
$\xi^{\mu}_{;i}=\xi^{\mu}_{;\gamma}\bar{\cal X}^{\gamma}_{,i}$ 
denotes  its projection over  $V_{n}$. 
The  curvatures of the higher dimensional space are distinguished
by a calligraphic  ${\cal R} $ .}. 
\begin{equation}
\bar{\cal X}^{\mu}_{,i}\bar{\cal X}^{\nu}_{,j}{\cal G}_{\mu\nu} =\bar{g}_{ij},\;
\bar{\cal X}^{\mu}_{,i}\eta^{\nu}_{A}{\cal G}_{\mu\nu}=0,\; 
{\eta}^{\mu}_{A}{\eta}^{\nu}_{B}{\cal G}_{\mu\nu}=g_{AB}   \label{eq:multi2}
\end{equation}
where we have denoted  by  ${\cal G}_{\mu\nu}$ the  metric of  $V_{D}$ in arbitrary coordinates  and  $g_{AB}$ denotes the components of the metric of  the complementary space $B_{N}$ in the basis $\{\eta_{A}\}$. 

One way to generate  a brane-world is   to  deform the background  space-time $\bar{V}_{n}$, in such a way that  it remains   compatible with the  gauge confinement and the  quantum fluctuations.   

The perturbations  of  an embedded geometry with respect to a small parameter $s$ along an  arbitrary transverse direction  $\zeta$ in  $V_{D}$  has been defined long ago \cite{Campbell,Nash,Geroch,Stewart,Gowdy}, starting with the  perturbation of the embedding map
 \begin{eqnarray}
{\cal Z}^{\mu}(x^{i},s) &= & \bar{{\cal X}}^{\mu}+s\pounds_{\zeta}
\bar{{\cal X}}^{\mu}) = \bar{\cal X}^{\mu} + s [\zeta,{\cal  X}]^{\mu} \label{eq:Defor1}
\end{eqnarray}
The presence of  a  component of  $\zeta$  tangent to $V_{n}$  is  a cause for concern because   it induces  coordinate gauges.  That is,  a perturbation could be altered  by a  mere coordinate transformation. In  general relativity, this problem  is  aggravated by the 
diffeomorphism  invariance of the theory.  There, the traditional   solutions consist  in  imposing specific  conditions on the  metric, tetrads  or  even on   the sources \cite{Chandra,Bardeen}. Another  solution,  consists in choosing a  hypersurface orthogonal perturbation (or equivalently, using the ADM language, to  eliminate  the  shift function), with the obvious limitations  resulting from the  use of  special coordinates \cite{Hojman}. 

In the case of brane-worlds, the  extra dimensions  do not  share the same  diffeomorphism invariance of  $V_{n}$.  Consequently,  the  limitations imposed by  the choice of  a  hypersurface  orthogonal 
vector  does not apply.  An additional  simplification is obtained by  taking the  norm  of  this vector  to be  $\pm 1$. This is  equivalent to normalize the   lapse function  to one  in the ADM case,  and using the  extra  coordinates  $s^{A}$  to play the role of the  "lapses".

With these precautions,  the  perturbations  of the  embedding map
 along  an  orthogonal direction $\bar{\eta}_{A}$,  for some fixed value of  $A$,  gives
\begin{eqnarray}
{\cal Z}^{\mu}_{,i}(x,s^{A}) =\bar{{\cal X}}^{\mu}_{,i}(x)  +s^{A}\bar{\eta}^{\mu}_{A,i}(x).
\label{eq:Zi}
\end{eqnarray}
On the other hand, since  $\bar{\eta}_{A}(x^{i})$ are independent vectors, depending only of $x^{i}$,  they  remain unperturbed
\begin{eqnarray}
\eta^{\mu}_{A}(x^{i})  =  \bar{\eta}_{A}^{\mu} +s^{B}[\bar{\eta}_{B},\bar{\eta}_{A}]^{\mu}=\bar{\eta}^{\mu}_{A} \label{eq:eta}
\end{eqnarray}  

The metric of the perturbed manifold  may be written in generic coordinates as  
\begin{equation}
g_{ij} =\bar{g}_{ij} + f_{ij}(x^{i},s^{A}) \label{eq:metric}
\end{equation}
Taking  $s^{A}$  small  as compared to one,  the Taylor expansion of  $f_{ij}$ in terms of  $s^{A}$,   gives the  four-dimensional gravitational field in the vicinity of the brane-world. For a  specified direction $\eta_{A}$, the linear perturbation assumes the form
\[
g_{ij} =\bar{g}_{ij} +s^{A}\gamma_{ijA}(x^{i})
\]

Applying   the de Donder gauge condition to the linearized Einstein's equations,    we obtain the  homogeneous gravitational wave equations  with respect to  the extra dimension  $\eta_{A}$. For  a  vacuum space-time   the  resulting wave equation reads as
\begin{equation}
\Box^{ij}{}_{k\ell}\Psi_{ij}^{A}(x,s) =0 \label{eq:deRahm}
\end{equation}
where  $\Psi_{ij}^{A} =\gamma_{ij}- 1/2\gamma \bar{g}_{ij}$,  $\gamma= \bar{g}^{mn}\gamma_{mn}$  and  where
we have denoted  the  generalized topological Dalambertian (de Rahm)   wave operator by
\[
\Box^{ij}{}_{k\ell}=\bar{g}^{ij}\nabla_{k}\nabla_{\ell} +2\bar{R}^{i\;\; \;j}_{\, k\ell} + 2\bar{R}^{i\;}_{ (k}\delta^{\; j}_{\ell)}
\]
We interpret  the solutions of   \rf{deRahm}  as  describing  gravitational waves over  the  curved background  $\bar{V}_{n}$,   in response to the quantum  fluctuations of the brane-world.
As  such, these gravitational  waves  belong to  the so-called  high frequency limit,   with a wavelength which is small as compared with the characteristic length of the background geometry  \cite{Brill,Isaacson,Zakharov}. To  define   this  property  we may  use   the Gaussian reference frame  based on the  perturbed  submanifold and the normal $\eta_{A}$. The  embedding equations for the perturbed geometry  are
 \begin{equation}
{\cal Z}^{\mu}_{,i}{\cal Z}^{\nu}_{,j}{\cal G}_{\mu\nu} =g_{ij},\;
{\cal Z}^{\mu}_{,i}\eta^{\nu}_{A}{\cal G}_{\mu\nu}=g_{iA},\; 
{\eta}^{\mu}_{A}{\eta}^{\nu}_{B}{\cal
G}_{\mu\nu}=g_{AB}   \label{eq:multi}
\end{equation}
where  now   we have   the mixed  metric  components   $g_{iA}=s^{M}A_{iMA}$, with 
\begin{equation}
A_{iAB}  =\eta^{\mu}_{B;i}\eta^{\nu}_{A}{\cal
G}_{\mu\nu}= \bar{\eta}^{\mu}_{B;i}\bar{\eta}^{\nu}_{A}{\cal G}_{\mu\nu}
=\bar{A}_{iAB} \label{eq:AiAB}
\end{equation}
The  extrinsic curvatures are 
\begin{equation}
k_{ijA}=-{\cal Z}^{\mu}_{,i}{\eta}^{\nu}_{A;j}{\cal G}_{\mu\nu} \label{eq:KijA}
\end{equation}
Replacing \rf{Zi} in  \rf{multi}, we obtain
the   perturbation of the metric 
\begin{eqnarray}
g_{ij} = \bar{g}_{ij}  -2s^{A}\bar{\kappa}_{ijA}
&+& s^{A}s^{B}(\bar{g}^{mn}\bar{\kappa}_{imA}\bar{\kappa}_{jnB} \nonumber\\
&+&  g^{MN} A_{iMA}A_{jNB}) \label{eq:pertu}
\end{eqnarray}
and the  perturbed extrinsic curvature in the same Gaussian frame 
\[
\kappa_{ijA} =\bar{\kappa}_{ijA} -
s^{B}(\bar{g}^{mn}\bar{\kappa}_{miA}\bar{\kappa}_{jnB} +g^{MN}{A}_{iMA}A_{jNB}) 
\]   
Equations  \rf{AiAB}, \rf{KijA} and \rf{pertu}  give the    evolution of the three basic geometrical attributes of the  perturbed geometry. Comparing  \rf{KijA} and the derivative of  \rf{pertu} we obtain   
\begin{equation}
\frac{\partial g_{ij}}{\partial s^{A}}=-2\kappa_{ijA} \label{eq:YORKG}
\end{equation}
which   generalizes  York's  relation  used in the ADM formalism.  Similar expressions were
applied to  geometric perturbations in \cite{Campbell,Nash}.  

The curvature radii  of the background $\bar{V}_{n}$ are   defined as  the   solutions of the homogeneous  equation
\begin{equation}
(\bar{g}_{ìj} -s^{A}k_{ijA})dx^{i}= 0  \label{eq:radius1},  \;\;  A \mbox{ fixed}.
\end{equation}
From $det (\bar{g}_{ìj} -s^{A}k_{ijA})=0$  we obtain  $n\times N$   distinct  solutions, 
$s=\rho^{A}_{i}$,  one for  each principal direction $dx^{i}$ and for each normal $\eta_{A}$. 
These are  local invariant properties  of  $\bar{V}_{n}$ and as such they are independent of the chosen Gaussian system \cite{Eisenhart}.  Since  \rf{pertu} can be written as   
\[
g_{ij} =\bar{g}^{mn}(\bar{g}_{im}-s^{A}\bar{k}_{imA})(\bar{g}_{jn}-s^{B}\bar{k}_{jnB})  + g^{MN}A_{iMA}A_{jNB}
\]
it  follows that  the  components
\begin{equation}
\tilde{g}_{ij} =\bar{g}^{mn}(\bar{g}_{im}-s^{A}\bar{k}_{imA})(\bar{g}_{jn}-s^{B}\bar{k}_{jnB})  \label{eq:tildeg}
\end{equation}
becomes singular  at the curvature centers.  Therefore,  the fluctuations of the brane-world
 which  are compatible with the regularity of the embedding should not reach  these points.
Since the smaller  solutions $\rho^{A}_{i}$  contribute more significantly  to the overall  curvature of  $\bar{V}_{n}$,  the  characteristic length of the background geometry  suitable for  comparing with the  wavelength  is  
\begin{equation}
\frac{1}{\bar{\rho}}= \sqrt{\bar{g}^{ij}g_{AB}\frac{1}{\bar{\rho}^{i}_{A}}\frac{1}{\bar{\rho}^{j}_{B}}}
\label{eq:RHO}
\end{equation}
which represents  a classical  limitation  of the perturbations.

\section{Dynamics}

 To   obtain the dynamical properties of the fluctuations  which are compatible with the embedding,  we apply the integrability conditions   for the embedding,  the Gauss, Codazzi and Ricci equations respectively
\begin{eqnarray}
R_{ijkl}& =& 2g^{MN}\kappa_{i[kM}\kappa_{jl]N} +{\cal R}_{\mu\nu\rho\sigma}{\cal Z}^{\mu}_{,i}
{\cal Z}^{\nu}_{,j}{\cal Z}^{\rho}_{,k}{\cal Z}^{\sigma}_{,l}\nonumber \\
\kappa_{i[jA,k]}& =& g^{MN}A_{[kMA}\kappa_{ij]N} +{\cal R}_{\mu\nu\rho\sigma}
{\cal Z}^{\mu}_{,i} \eta^{\nu}{\cal Z}^{\rho}_{,j}{\cal Z}^{\sigma}_{,k}\label{eq:GCR}\\
2A_{[jAB;k]} &=& - 2g^{MN}A_{[jMA}A_{k]NB} \nonumber \\
 &-& g^{mn}\kappa_{[jmA}\kappa_{k]nB} - {\cal R}_{\mu\nu\rho\sigma}{\cal
Z}^{\rho}_{,j} {\cal Z}^{\sigma}_{,k}\eta^{\nu}_{A}\eta^{\mu}_{B} \nonumber
\end{eqnarray}
These equations  represent the  conditions for a perturbation of  $\bar{V}_{n}$   to be 
stable as   a    submanifold  (and in particular a  brane-world) \cite{Eisenhart}. 

The relevance of   the  above equations  to the stability of brane-worlds  has been  pointed out in some particular situations, using the  conformal  flatness properties \cite{Tanaka,Maeda}.  In this section we show that  that  the  implications of   \rf{GCR}  to the   dynamics of  brane-worlds is   quite general, independently of  models or any  additional assumption on the geometries of  $V_{D}$ and  $V_{n}$.  For this, purpose, consider  the expression  $\xi^{\mu\nu}=g^{ij}Z^{\mu}_{,i}Z^{\nu}_{,j}$. From  \rf{multi}  it follows that 
\[
\xi^{\mu\nu}{\cal G}_{\mu\nu}= n\;\; \mbox{and}\;\;  \xi^{\mu\nu}\eta^{\alpha}_{A}{\cal G}_{\mu\alpha} =0
\] 
so that  $\xi^{\mu\nu}$  cannot be proportional  to  ${\cal G}^{\mu\nu}$.  Writing $\xi^{\mu\nu}={{\cal G}}^{\mu\nu} 
+\zeta^{\mu\nu}$,  we find that $\zeta^{\mu\nu}$ must satisfy 
\[
{\cal G}_{\mu\nu}\zeta^{\mu\nu}=-N\;\; \mbox{ and}\;\;\, 
\zeta_{\mu\nu}\eta^{\mu}_{A}\eta^{\nu}_{B}=-g_{AB}
\]
 The  solution of   these  algebraic  equations,   compatible with  \rf{multi}  is 
$\zeta^{\mu\nu} =g^{AB}\eta^{\mu}_{A}\eta^{\nu}_{B}$, so that
\begin{equation}
g^{ij}{\cal Z}^{\mu}_{,i}{\cal Z}^{\nu}_{,j}  ={\cal G}^{\mu\nu}
-g^{AB}\eta^{\mu}_{A}\eta^{\nu}_{B}  \label{eq:INV1}
\end{equation}
Applying this  result in the contractions of  the first equation  \rf{GCR}, we obtain  the Ricci scalar for  the perturbed geometry
\begin{eqnarray}
R  &=& (\kappa^{2} -h^{2})  +{\cal R}  -2g^{MN}{\cal R}_{\mu\nu} \eta^{\mu}_{M}\eta^{\nu}_{N} \label{eq:RICCI}\\
& - & g^{AB}g^{MN}{\cal R}_{\mu\nu\rho\sigma}\eta^{\mu}_{A}\eta^{\sigma}_{B}\eta^{\nu}_{M}
\eta^{\rho}_{N} \nonumber       
\end{eqnarray}
where  we have denoted $\kappa^{2}=\kappa_{ijA}\kappa^{ijA}$ and the  mean  curvatures
$h_{A}=g^{ij}\kappa_{ijA}$   with norm   $h^{2} =g^{AB}h_{A}h_{B}$.
Using the  Gaussian  frame we see that  the last term vanishes and  that
\[
g^{AB}{\cal R}_{\mu\nu}\eta^{\mu}_{A}\eta^{\nu}_{B}  =-g^{AB}\frac{\partial
h_{A}}{\partial s^{B}} +\kappa^{2}
\]
Therefore,  \rf{RICCI}reduces to
\[
 R ={\cal R} - (\kappa^{2} + h^{2})  -2 g^{AB}\frac{\partial h_{A}}{\partial s^{B}}
\]
To obtain  the  canonical structure  associated with  the perturbations  we  may write the Einstein-Hilbert Lagrangian for  $V_{D}$, after discarding the derivative terms ${\partial h_{A}}/{\partial s^{B}}$  as surface terms\footnote{To allow for different signatures,  we  have denoted  ${\cal G}= {det({\cal G}_{\alpha\beta})}$  module  the signature  of the  extra dimensions.}.
\begin{equation}
{\cal  L}(g,   g_{,i}, g_{,A})={\cal R}\sqrt{{\cal G}}=  \left[ R  + (\kappa^{2} + h^{2})\right] \sqrt{{\cal G}}\label{eq:EH}
\end{equation}

The components of the momentum  canonically conjugated to ${\cal G}_{\alpha\beta}$ relative
to the normal direction $\eta_{A} $ are
\[
p^{\alpha\beta}_{( A)} =\frac{\partial {\cal L}}{     \partial \left(
 \frac{ \partial {\cal G}_{\alpha\beta} }{\partial s^{A}}   \right)     }
\]
In particular, using  \rf{YORKG} we obtain the  tangent components
\begin{equation}
p^{ij}_{(A)}=-(\kappa^{ij}_{A} +h_{A}g^{ij})\sqrt{{\cal G}} \label{eq:PijA}
\end{equation}
which describe the momentum  the metric geometry of  $V_{n}$,  when propagated  along the  extra dimensions. 

On the other hand, the perturbation does not prescribe  the evolution  of   ${\cal G}_{iA}$ and ${\cal G}_{AB}$. The corresponding momenta  are be set  as   constraints:
\begin{eqnarray}
p^{iA}_{(B)}  &= &  -2\frac{\partial{\cal
R}_{\alpha\beta}\eta^{\alpha}\eta^{\beta} }{\partial \frac{\partial{\cal
G}_{iA}}{\partial s^{B} }} \sqrt{{\cal G}}=0  
\label{eq:DEFCON1},\\
p^{AB}_{(C)} & = & -2
\frac{\partial {\cal R}_{\alpha\beta}\eta^{\alpha}_{A}\eta^{\beta}_{B}}
{\partial \frac{\partial {\cal G}_{AB}}{\partial s^{C}}}\sqrt{{\cal G}}=0 \label{eq:DEFCON2}
\end{eqnarray}

The constraint  \rf{DEFCON1}   correspond to our previous choice of  zero shift function, while
  \rf{DEFCON2} corresponds to the   normalization of the lapse. 

The Hamiltonian  for  each single  direction  $\eta^{A}$,  is defined by the  Legendre transformation 
\begin{eqnarray}
{\cal H}_{A}(g,p) &=&   p^{ij}_{(A)}g_{ij,A}-{\cal L} =\nonumber \\
 &-& R\sqrt{{\cal G}}  -\frac{1}{{\cal G}}\left(\frac{p_{A}^{2}}{n+1}
- p_{ij (A)}p^{ij}_{(A)} \right) \label{eq:HG}
\end{eqnarray}
 leading to   Hamilton's equations,
\begin{eqnarray}
\frac{d g_{ij}}{d s^{A}} & =& \frac{\delta {\cal H}_{A}}{\delta p^{ij (A)}}=
\frac{-2}{\sqrt{\cal G}} \left(\frac{g_{ij}p_{A}}{n+1}-p_{ij(A)}\right),\vspace{3mm}\label{eq:DOTG}\\
\frac{d  p^{ij}_{(A)} }{ d s^{A} } & = & -\frac{\delta {\cal H}_{A}}{\delta g_{ij}}= G_{ij}\sqrt{g\varepsilon} + \frac{1}{\sqrt{g\varepsilon} }
\left[ \frac{2p_{A}p_{ij(A)}}{n+1} \right. \nonumber\\
&+& \left. \frac{1}{2}\left(
\frac{p_{A}^{2}}{n+1} +p_{ij(A)}p^{ij}_{(A)} \right) g_{ij}\right] \label{eq:DOTP}
\end{eqnarray}
The first of these equations  coincides with   \rf{YORKG}
 expressed in terms of  $p_{ij(A)}$. 
 The second equation  gives  the  evolution of the  extrinsic curvature of the perturbation  in terms of  momentum.  Consequently, the stability  conditions  for the  perturbations, represented by \rf{GCR} are  consistent  with  the Einstein-Hilbert  dynamics for  $V_{D}$.

\section{Hypersurface Brane-worlds}

When   $D= n+1$,  the brane-worlds are  hypersurfaces of  $V_{D}$. All expressions can be derived from the  general case by setting $A,B\cdots  =n+1$ and $g_{AB}=g_{n+1\,n+1}=\varepsilon =\pm1$.
Since in this case the "twisting vector" $A_{iAB}$  vanishes,  the integrability conditions lose  Ricci's equation.  The remaining  Gauss-Codazzi equations may be applied to
derive the dynamics of  the  hypersurface for  a general  $V_{(n+1)}$.
Expression \rf{INV1} in this case  becomes
\begin{equation}
g^{ij}{\cal Z}^{\mu}_{,i}{\cal Z}^{\nu}_{,j}  ={\cal G}^{\mu\nu}
-\frac{1}{\varepsilon}\eta^{\mu}\eta^{\nu}  \label{eq:INV}
\end{equation}
As before, replacing in Gauss'  equation, after removing the total derivative terms  the  Lagrangian  \rf{EH} reduces to  
\begin{equation}
{\cal L}=\left[ {R} +
\frac{1}{\varepsilon}(\kappa^{2}+h^{2})\right]  \sqrt{{\cal G}}\label{eq:L} 
\end{equation}
where now  we have denoted $h=g^{ij}k_{ij}$  and    $k^{2}=k^{ij}k_{ij}$.

The  momentum canonically conjugated to the metric ${\cal G}_{\alpha\beta}$,  with respect to the  perturbation parameter  $s$ is (here the dot means derivation with respect to $s$) 
$p^{\alpha\beta}={\partial{\cal L}}/{\partial (\dot{\cal G}_{\alpha\beta}) } $,
with components
\begin{eqnarray*}
p^{ij}& =& \frac{-1}{\varepsilon}( k^{ij}+hg^{ij} )\sqrt{{\cal G}} \label{eq:piij}\\
p^{i\, n+1} & = &  -2\frac{\partial{\cal
R}_{\alpha\beta}\eta^{\alpha}\eta^{\beta} }{\partial \dot{{\cal
G}}_{i\,n+1}}\sqrt{{\cal G}}=0  
\label{eq:DEFCON1},\\
p^{n+1\,n+1} & = & -2\frac{\partial {\cal R}_{\alpha\beta}\eta^{\alpha}\eta^{\beta} }
{\partial {\dot{\cal G}}_{n+1\,n+1}}\sqrt{{\cal G}}=0 \label{eq:DEFCON2}
\end{eqnarray*}
The  Hamiltonian  corresponding to the one parameter  perturbation is
\begin{eqnarray}
{\cal H}=p^{\alpha\beta}\dot{\cal G}_{\alpha\beta}-{\cal L}  &=&\nonumber\\
-R\sqrt{{\cal G}} &-&\frac{\varepsilon}{{\cal G}}\left( \frac{p^{2}}{n+1} -p_{ij}p^{ij}
\right) \sqrt{{\cal G}}
\label{eq:H}
\end{eqnarray}
where    $p={\cal G}_{\alpha\beta}p^{\alpha\beta}$. This   describes the  same   perturbations of an arbitrary metric  with   some limitations  to be noted:

Since $A_{iAB}=0$,  the   perturbed metric in the Gaussian frame becomes simply
\[
g_{ij}=\tilde{g}_{ij}=\bar{g}^{mn}(\bar{g}_{im}-sk_{im})(\bar{g}_{jn}-sk_{jn})
\]
which is  singular at the  curvature centers of   $V_{n+1}$  defined by $det(\bar{g}_{ij} -sk_{ij})=0$.  The solutions of this equation are  the  curvature radii
$\rho_{i}$, one for each principal direction  $dx^{i}$.  The  characteristic length  has  
more significant  contributions  from the smaller $\rho_{i}$, in accordance with  \rf{RHO},
\[
\frac{1}{\bar{\rho}} = \sum\sqrt{\varepsilon \frac{\bar{g}_{ij}}{\bar\rho_{i}\bar\rho_{j} }}
\]
This is again  an invariant property  of the  embedded background geometry. As it was   already commented, the  properties of  this  characteristic length  have implications on the 
perturbations  of  brane-world geometry. In particular, a  known result states that if  $V_{n}$  has more than two finite curvature radii  $\rho_{i}$, then  the  hypersurface  becomes indeformable \cite{Eisenhart}. 

A particularly interesting  example is given by a  constant curvature   space $V_{(n+1)}$. More specifically,  consider  the (non-compact) anti-De Sitter space $AdS_{(n+1)}$,  with $\epsilon =-1$:
\begin{equation}
{\cal R}_{\mu\nu\rho\sigma}=-\Lambda_{*}({\cal G}_{\mu\rho}{\cal G}_{\nu\sigma}-{\cal G}_{\mu\sigma}{\cal G}_{\nu\rho}) \label{eq:ADS}
\end{equation}
We obtain,  ${\cal R}_{\mu\nu}= n \Lambda_{*}{\cal G}_{\mu\nu}$, so that
the constraints \rf{DEFCON1}, \rf{DEFCON2} become identities and   $\Lambda_{*}$ can be interpreted as  a  bulk cosmological constant.  

In this particular case, the integrability conditions  are considerably simpler:
\begin{eqnarray*}
R_{ijkl}& =& \frac{2}{\varepsilon}  \kappa_{i[k}\kappa_{l]j} -
\Lambda_{*}(g_{ik}g_{jl}-g_{il}g_{jk})\\
\kappa_{i[j;k]}& =& 0
\end{eqnarray*}
where we notice that  the  Riemann tensor of  $V_{(n+1)}$ is  entirely projected into the  $n$-dimensional hypersurface. Therefore,  it is more natural to  derive Einstein's equations
for  $V_{n}$. Using \rf{INV},  we obtain
\begin{equation}
G_{ij}= \frac{1}{\varepsilon} t_{ij}(k)  +\Lambda_{*}(\frac{n}{2}-1)(n-1)g_{ij}  =8\pi G T_{ij}\label{eq:EC}
\end{equation}
where  we  have denoted     
\[
t_{ij}(k) =k^{m}_{i}k_{mj} -hk_{ij} -\frac{1}{2}(k^{2}-h^{2})g_{ij}
\]
and  $T_{ij}$  denotes the  energy-momentum tensor of the four-dimensional sources. 
Since the last equality in \rf{EC} is  not a differentiable  equation on  $g_{ij}$,   it implies  that  this energy-momentum tensor   becomes  algebraically related to  the  extrinsic  curvature  $k_{ij}$ which represents  a  serious limitation for the  brane-world.
In this respect,  it has been   noted that  when  $\Lambda_{*} <0$, the  null energy condition
for  $T_{ij}$  is not compatible  with the  embedding  \cite{Manheim:1,Barcelo}.

We will see in section VI  that the condition $A_{iAB}=0$ means  that  the  gauge structure  does not  arise from  the  integrability conditions. Consequently it has  to be  imposed over the geometrical structure,  requiring additional conditions  to  ensure the stability of  the hypersurface brane-worlds. 

\section{Quantum states}
 
Klein's compactification of  the extra dimension was  introduced to make  Kaluza's theory  compatible with  quantum mechanics. Specifically, the normal modes of the harmonic  expansion  with respect to the  internal parameters  (of Planck's length size) were  set in correspondence with the quantum modes \cite{Klein}.  This eventually  led to a major   problem,  namely the inability to generate light chiral fermions at  the   electroweak   sector of the theory. More specifically,  the   strong curvature of the internal space  contributed to large  mass  fermion  states, which would  necessarily be observable at the electroweak scale. 

On the other hand, in brane-worlds  the   extra dimensions  are macroscopic, so that
the harmonic expansion of physical fields over  $V_{D}$ would not   lead to the  correct quantum phenomenology. In its place, we look at the linear  gravitational wave
equation \rf{deRahm} in the  de Donder gauge.  However, two points must be observed:
Firstly, the quantum  fluctuations  of the geometry should be independent of the classical approximation order on  $s^{A}$.  Furthermore, the classical waves correspond to the  quantum  fluctuations of the geometry. This  suggests that the  definition of the quantum states  should precede the linear approximation of the metric.

One way to  define the quantum states  relative to the extra dimensions is to
use the canonical formalism  associated with the perturbations.
 In fact, the Poisson bracket  structure  for  each  extra dimension  $\eta_{A}$  defined  by
\[
[{\cal F},{\cal H}_{A}]=\frac{\delta {\cal F}}{\delta g_{ij}}\frac{\delta {\cal H_{A}}}{\delta  p^{ij(A)}}-\frac{\delta {\cal F}}{\delta p^{ij(A)}}\frac{\delta {\cal H_{A}}}{\delta  g_{ij}}
\]
can be derived  consistently with \rf{HG}, \rf{DOTG}  and \rf{DOTP}, with the  commutator  between two  independent  perturbations    given by  $[{\cal H}_{A},{\cal H}_{B}]$. Therefore the   quantum  fluctuations  may be defined for  each  independent  extra dimension $\eta_{A}$,  and  the superposition principle  applied afterwards. 

As one  example,  consider  Schr\"odinger's  equation
\begin{equation}
-i\hbar \frac{ d \Psi_{ij(A)} }{ d s^{A}} =\hat{\cal H}_{A}\Psi_{ij(A)}
\label{eq:SCA}
\end{equation}
where  the operator $\hat{\cal H}_{A}$ is  constructed with the perturbation Hamiltonian 
\rf{HG}. 
The probability  of  a  brane-world to be  in a state   $\Psi_{ij(A)}$ is  given by the Hilbert norm
\[
<\Psi_{ij(A)},\Psi_{ij(A)}>=\int \Psi_{ij(A)}^{\dagger}\Psi_{ij(A)}\,dv
\]
where  the integral extends over a volume  in $V_{D}$  with a base on a  compact region 
of the  background and  a finite extension of the extra coordinates  $s^{A}$.
 Considering two  independent  directions  $\eta_{A}$ and  $\eta_{B}$,
the  transition probability between the corresponding states is  given by  the integral
$<\Psi_{ij\,(A)},\Psi_{k\ell\,(B)}>$.

Topological changes  are expected  to occur in  any  quantum theory of  space-times \cite{Hawking,Balachandra,Dowker,Sorkin}. Therefore,  we cannot use a fixed product topology for  $V_{D}$,  as  in \cite{Arkani:1}. 
With  multiple  evolution parameters  we  may have a  more complex  topological  variation than  those expected in a single time theories.
For example,  if   $\eta_{A}$ and  $\eta_{B}$ are  both space-like, then   $<\Psi_{ij\, (A)},\Psi_{k\ell\, (B)}>$   corresponds
to  a  space-like  handle.  
On the other hand, if  $\eta_{A}$ and  $\eta_{B}$  have  both time-like signatures, then 
the classical limit  of  the transition probability  corresponds to a classical loop involving two internal time parameters,  suggesting a multidimensional time machine.  Finally, if  $\eta_{A}$ and  $\eta_{B}$  have different signatures, then the transition probability  $<\Psi_{ij\, (A)},\Psi_{k\ell\, (B)}>$  corresponds to a signature change.

 An  example of the last case is given by the  Kruskal brane-world  regarded as  a perturbation of the embedded Schwarzschild  space-time, in such a way that the latter becomes geodesically complete. These spaces  are both  embedded in  six dimensional flat spaces,   with   signatures   $(5,1)$ and $(4,2)$ respectively \cite{Fronsdal:1}. 
The Schwarzschild space-time  is a subset of  Kruskal space-time,  but they do not
belong  to the same fixed embedding space.  However, they may be considered as  classical  limits of  distinct  quantum  states of the dynamically embedded Kruskal brane-world, with a signature transition at the horizon.

As  a  final remark  we add that  due to the extrinsic nature of  the parameters $s^{A}$,  equation  \rf{SCA} may be  understood  as  first  quantization of the  brane-world geometry. However, it does not exclude the  quantization of the  metric as  an  effective field  theory  in four dimensions,  taking    for example  $V_{D}$ as the   space of all  deformed  metrics  \cite{Ashtekar,DeWitt}.

\section{Confinement}

The confinement  hypothesis for gauge fields implies that  regardless of the  electromagnetic, weak and  strong  interactions taking place inside  the  brane-world,  its  differentiable structure
 remains intact.  Since,  those interactions  are concomitant  with the  quantum fluctuations of the geometry, it follows that   equations  \rf{GCR} must also be compatible  with the  confinement of the gauge fields.
 
The  basic  field variables in \rf{GCR} are  $g_{ij}$, $\kappa_{ijA}$ and  $A_{iAB}$. The first two  vary with the perturbation and they are related by \rf{YORKG}. On the other hand, from   \rf{AiAB} it follows that   $A_{iAB}$  remain  confined in the sense that  they remain unchanged,   independently of  the quantum  fluctuations of the brane-world. 
 The  relevant but so far little explored fact, is that   those  components transform  as the components of a   gauge potential  under  the  group  of  isometries of the complementary  space  $B_{N}$. This  follows directly from  the  transformation of the  mixed component of the  metric tensor,
under a local  infinitesimal   coordinate  transformation of  $B_{N}$ 
\[
 s'^{A}= s^{A} +\xi^{A}\;\;\mbox{with}\;\;  \xi^{i}=0,\;\; \mbox{and}\;\;  \xi^{A} a =\Theta^{A}_{M}(x^{i})s^{M}
\]
where  $\Theta^{A}_{B}$  are the  infinitesimal parameters.
Denoting   generic coordinates  in  $V_{D}$ by $\{x^{\mu}\} =\{x^{i},s^{A}\}$, it follows that  
\[
g'_{iA} =g_{iA} + g_{i\mu}\xi^{\mu}_{,A} +g_{A\mu}\xi^{\mu}_{,i} +\xi^{\nu}\frac{\partial  g_{iA}}{\partial x^{\nu}} +0(\xi^{2})
\]
The transformation of  $A_{iAB}$  follows from
\[
A'_{iAB}=\frac{\partial  g'_{iA}}{\partial  s'^{B}}= \frac{\partial  g'_{iA}}{\partial  s^{B}}
- \xi^{\mu}_{,B}\frac{\partial g'_{iA}}{\partial x^{\mu}}
\]
Using  $\xi^{A}_{,B}=\Theta^{A}_{B}(x^{i})$, $\xi^{A}_{,i} =\Theta^{A}_{B}s^{B}$ and  $\xi^{A}_{BC}=0$  we obtain
\begin{equation}
A'_{iAB} =  A_{iAB} -2g^{MN}A_{iM[A}\Theta_{B]M} +g_{MB} \Theta^{M}_{A,i}
\end{equation}
which  shows that  indeed  $A_{iAB}$ transforms  as a gauge potential.
Once the group  of  isometries of $B_{N}$ has been characterized,  the  above transformation
may also be written in terms of its structure constants  \cite{Maia:3,Maia:4}.

To conclude  the  characterization of   $A_{iAB}$,  we notice that  the metric of  $V_{D}$   written in the Gaussian frame has separate has  components  
\begin{eqnarray}
g_{ij} &= &{\cal Z}^{\mu}_{i,}{\cal Z}^{\nu}_{,j}{\cal G}_{\mu\nu}=
\bar{g}_{ij} -2s^{A}\bar{k}_{ijA}\nonumber\\
&+& s^{A}s^{B}\left ( \bar{g}^{mn}\bar{k}_{imA}\bar{k}_{jnB}
+\bar{g}^{MN}A_{iMA}A_{jNB}\right )\nonumber\\ 
g_{iA} & = & {\cal Z}^{\mu}_{,i}\eta^{\nu}_{A}{\cal G}_{\mu\nu} =
s^{M}A_{iMA}\nonumber  \\
g_{AB} & =&\eta^{\mu}_{A}\eta^{\nu}_{B}{\cal G}_{AB} \nonumber 
\end{eqnarray}
or, in matrix notation,
\begin{equation}
{\cal G}_{\alpha\beta}=
\left( \matrix{ \tilde{g}_{ij}  + g^{MN}A_{iM}A_{jN}  &   A_{iA} \cr
                           A_{jB} &  g_{AB}}
\right) \label{eq:KK}
\end{equation}
where   $\tilde{g}_{ij}$ is given by \rf{tildeg} and
\begin{equation}
  A_{iA}  =  s^{M}A_{iMA} \label{eq:AiA}
\end{equation}
The  metric  \rf{KK} has the same appearance as the Kaluza-Klein metric ansatz, with the exception that  $\tilde{g}_{ij}$  is  not  the   background metric  but rather an untwisted  perturbation of it,  
given  by \rf{tildeg}.

The Einstein-Hilbert  Lagrangian  derived directly from \rf{KK} is
\begin{equation}
{\cal L } ={\cal R}\sqrt{\cal G}  ={R}\sqrt{\tilde{g}\epsilon}
+\frac{1}{4}tr {F}^{2}\sqrt{\tilde{g}\epsilon} 
 \label{eq:EYM}
\end{equation} 
where  we have  denoted $\epsilon= det{(g_{AB})}$,
  ${F}^{2}={F}_{ij}{F}^{ij}$ and  ${F}_{ij}=[D_{i},D_{j}]$, 
$D_{i}=\partial_{i} + A_{i}$.  Considering the gauge group as the group of isometries of  $B_{N}$, the  gauge  connection $A_{i}$  can be  expressed in the Killing basis    $\{K^{AB}\}$  of the corresponding Lie algebra as
\[
{A}_{i}=A_{iAB}K^{AB}
\]

From \rf{EYM}  we see that the  bulk gravitational field  described by  ${\cal G}_{\alpha\beta}$  decomposes in  a four dimensional gravitational  interaction represented  by  $\tilde{g}_{ij}$,  plus the gauge interactions represented by   $A_{iAB}$,  much in the  sense of  Kaluza-Klein theory.

\section{Conclusion}
We have  taken an approach to  brane-world  theory that is  independent of the  two
known models proposed in \cite{Arkani:1,Randall}. 
Starting  with the classic perturbation analysis of     submanifolds, 
the  geometric stability conditions \rf{GCR} for the  perturbation are assumed to hold throughout. The conclusion is that  those equations  in fact determine the  brane-world dynamics in the  general case.  

From the fact that the  internal parameters do not share the same  diffeomorphism invariance  of the gauge fields,  we have described a  non-constrained canonical structure which
 reproduces the classical perturbation equations.  The   perturbation Hamiltonians, one for each extra dimension,  is  compatible with  a canonical  quantization relative  to the internal parameters $s^{A}$. It was  also  noted that  the  quantum fluctuations in general  are not  compatible with the  adoption of a fixed topology for  $V_{D}$.
 
We  find it significant to  brane-worlds the fact that   when  $N\ge 2$,  the twisting vector  $A_{iAB}$,  has the structure of  a  confined gauge  field built in  \rf{GCR}.
 
It is  normally assumed that the  brane-world is  four-dimensional.  However, this assumption must be  consistent with the properties of the embedding and the  gauge group $G$.   Since the host space   $V_{D}$ may be  thought of as  populated by
 point particles to (n-1)-dimensional objects, the end  product  of their dynamics is the n-dimensional  brane-world regarded as a  dynamically  embedded space-time.
As such, the  differentiable  equations   \rf{GCR} assume  the role of   structure preserving equations for brane-worlds embedded
in a  $V_{D}$ with  $D\le n(n+3)/2$.  Setting  $D=n+N$, it follows that  $n^{2} +n -2N\ge 0$.  

For the  standard model gauge group $SU(3)\times SU(2)\times U(1)$ acting on
a seven dimensional projective space,  we  find    $n \ge 3.27$, meaning that the standard model  just fits in  four-dimensional  stable brane-worlds.
On the other hand,   using the  $SO(10)$  GUT  we obtain  exactly  $n=4$, suggesting  that  four dimensional brane-worlds  have  a  natural  existence in 
a particular fourteen dimensional  theory   with signature $(11,3)$  and  with   $SO(1O)$ as   the gauge group.    

Our final  remark  concerns  the   size  of the extra dimensions and the consequent  modification of Newton's law at small distances. 
From  \rf{tildeg}  it  follows that   the  perturbation becomes   singular  at the   curvature centers of the background  characterized by  $\rho^{A}_{i}$, so that  $s^{A}$ must remain in the open  intervals  $[0,\pm\rho^{A}_{i})$. This means that  as  long as the curvature radii   $\rho^{A}_{i}$  remain finite, the  volume  of the  complementary  space  available for  the graviton probes is  finite.  In this case, noting that the right hand side of  \rf{EH}, depends only on $x^{i}$,  we may evaluate the  action integral in that  region 
 \[
\int{\cal L}\sqrt{\cal G}d^{n+N}v= \int\left(\int[R  -(k^{2} +h^{2})]\sqrt{\tilde{g}}\sqrt{\epsilon}\,d^{n}v\right)d^{N}v
\]
which leads to an argument similar to that in \cite{Arkani:1}, but without assuming the product topology:
\[
\frac{1}{M_{\ast}^{2+N}}= \frac{1}{M_{pl}^{2}}(1 -K) {\cal V}
\]
where  we have denoted  $K= \int(k^{2} +h^{2})]\sqrt{\tilde{g}\epsilon}\,d^{n}v$,  $
{\cal V}$ is the  finite volume of the  region in the  complementary space and  $M_{\ast}$ is the  Planck's  mass in the bulk. 

 The  volume  ${\cal V}$ depends on the curvature of  $\bar{V}_{n}$ and  it is detailed by the  different  values of  $\rho^{A}_{i}$  as  given by  \rf{RHO}.
Therefore,  defining  an internal   spherical space  with  radius  $\bar{\rho}$,  we  have 
  ${\cal V}\approx  \bar{\rho}^{N}$, so that   
\[
\bar{\rho}^{N} \approx \frac{M_{pl}^{2}}{M_{\ast}^{2+N}}\frac{1}{(1 -K)},\;\; K\neq 1
\]
This  gives  the same  estimates as in  \cite{Arkani:1}   when $K$ is  small as  compared to one.  The case  $N=1$   has been  ruled out in our analysis  because of the limitations imposed on the  brane-world  fluctuations.  The actual distances  probed by gravitons  depend on  further  analysis  of the quantum states.

\acknowledgements{ The authors  wish to acknowledge  the stimulating discussions on the subject with Drs. P.  Caldas} and   Vanda Silveira.

\end{document}